# Shear induced normal stress differences in aqueous foams

Vincent Labiausse, Reinhard Höhler, and Sylvie Cohen-Addad

Université de Marne-la-Vallée
Laboratoire de Physique des Matériaux Divisés et des Interfaces, UMR 8108,
5 Boulevard Descartes, 77 454 Marne-la-Vallée cedex 2, France

A finite simple shear deformation of an elastic solid induces unequal normal stresses. This nonlinear phenomenon, known as the Poynting effect, is governed by a universal relation between shear strain and first normal stresses difference, valid for non-dissipative elastic materials. We provide the first experimental evidence that an analog of the Poynting effect exists in aqueous foams where besides the elastic stress, there are significant viscous or plastic stresses. These results are interpreted in the framework of a constitutive model, derived from a physical description of foam rheology.

## 1. Introduction

Subjected to small strains, aqueous foams show solid-like viscoelastic behavior. For increasing strains, nonlinear viscoelastic response sets in and finally, beyond a yield strain, a cross-over to non Newtonian liquid-like behavior is observed. On the scale of the bubbles, the cross-over is governed by strain induced local bubble rearrangements. This rheological behavior gives rise to many applications, such as oil production, mineral separation processes or food production (Höhler & Cohen-Addad 2005; Khan & Prud'homme 1996; Weaire & Hutzler 1999). In this context, as well as from a fundamental point of view, it is important to develop a constitutive rheological law, predicting the relation between macroscopic stress, strain and strain history. Such a law should be based on an understanding of the underlying physical processes at the scale of the gas-liquid interfaces, the bubbles, and possibly on a mesoscopic scale. In the linear regime, the well-known Princen law predicts the static shear modulus as a function of liquid fraction, average bubble size and surface tension (Princen & Kiss 1986; Saint-Jalmes & Durian 1999). However, truly static elastic behavior does not exist in real foams who are intrinsically non-equilibrium systems (Cohen-Addad et al 2004; Gopal & Durian 2003). Even if drainage and bubble coalescence are suppressed, there is a time evolution due to coarsening: The Laplace pressure differences between neighboring bubbles drive a diffusive gas transfer through the liquid films, on the average from smaller towards larger bubbles. As a consequence, local strains build up and are released by intermittent rearrangements. They reduce the elastic energy of the bubble packing and, on the average, they relax the macroscopically applied strain. As a result, coarsening foams show a creep flow that linearly progresses with time and whose speed scales as the rate of bubble rearrangements (Cohen-Addad et al 2004; Marze et al 2005; Vincent-Bonnieu et al 2006). On very long time scales, they may therefore be regarded as viscoelastic Maxwell liquids rather than solids. An additional slow relaxation mechanism in foams is related to the intrinsic viscous behavior of the surfactant layer adsorbed to the gas-liquid interfaces (Cohen-Addad et al 2004). On short time scales, typically corresponding to frequencies above $1 - 10$ Hz, a viscoelastic behavior asymptotically described by the relation $G^* \propto (i\omega)^{1/2}$ has been observed in foams and concentrated emulsions (Cohen-Addad et al 1998; Gopal & Durian 2003; Liu et al 1996). In this paper, we will focus on the low frequency response.

When an elastic material is subjected to shear strains that are large but insufficient to induce significant plastic flow or yielding, the induced shear stress is accompanied by unequal normal stresses, a phenomenon known as the Poynting effect (Mal & Singh 1991; Poynting 1909; Rivlin 1953). Since aqueous foams can be considered as incompressible, in the sense that their compression modulus is orders of magnitude larger than their shear modulus, the ambient hydrostatic pressure is not a significant parameter and the Poynting effect is described in terms of the normal stress *differences* $N_1$ and $N_2$ whose definition is illustrated in Fig. 1. To validate nonlinear viscoelastic constitutive laws, measuring these shear induced normal stress differences is of fundamental interest. For isotropic materials whose stress is a function of the instantaneous strain only, very general arguments recalled in section 2.1 show that the first normal stress difference and the shear strain $\gamma$ are related by the following law, found by Rivlin in his analysis of Poynting's pioneering experiments (Poynting 1909; Rivlin 1953):

$$N_1 = \sigma_{12}\gamma \qquad (1)$$

In the following, we will refer to Eq. (1) as the Poynting relation. Quasi-static numerical simulations of sheared disordered aqueous foams, based on the principle that foam equilibrium structures must be configurations of minimal gas liquid interface area, are in agreement with this relation, in the elastic regime where strain induced bubble rearrangements are negligible (Kraynik & Reinelt 2004; Kraynik et al 2000). Moreover, analytical models relying on a statistical description of the foam structure (Doi & Ohta 1991; Höhler et al 2004; Larson 1997) as well as isotropically averaged calculations for ordered foam structures (Reinelt 1993) also agree with this result. However, the theoretical argument demonstrating the Poynting relation no longer holds in the presence of strong viscous dissipation which is not considered in the mentioned theoretical studies. Moreover, it is questionable whether real foams are indeed statistically isotropic materials, as required for the Poynting relation to hold and as assumed in the mentioned models: Since foams are yield stress fluids, they conserve a rheological memory of their flow history in the form of anisotropic trapped stresses. It has been shown experimentally that even the linear shear modulus of foam is a function of flow history (Höhler et al 1999). For these reasons, it is of interest to confront the Poynting relation, describing the nonlinear rheological response, to experimental evidence obtained for aqueous foams.

We are not aware of any previous successful experimental studies of the first normal stress difference in the solid-like regime for viscoelastic yield stress fluids such as foams and concentrated emulsions. It has been reported in the literature that attempts to measure $N_1$ for foams in the solid-like regime failed, due to the uncontrolled trapped plastic strains, created during the sample preparation (Khan et al 1988). In this paper, we present new experimental and theoretical approaches, allowing to study the first normal stress difference response of yield strain fluids and we apply them to dry aqueous foam. Let us note that any uniaxial strain also induces elastic normal stress differences. However, this effect which has recently been observed in 2D foam (Asipauskas et al 2003) is to leading order *linear* and therefore differs qualitatively from the *nonlinear* elastic effect whose quantitative experimental observation in foams is to our knowledge reported for the first time in the present paper.

## 2. Theory

To make this paper self contained, we first recall some fundamental results of nonlinear elasticity and we outline the proof of the Poynting relation that can be found in the literature (Larson 1988). In section 2.2, we use a physically motivated static constitutive law to analyze the influence of trapped strains on the nonlinear rheological response of foams, and we show that for these materials, normal stress measurements using oscillating imposed strain are much better suited for probing the intrinsic constitutive law than more conventional step strain experiments. In section 2.3, we present a simple constitutive relation which allows studying viscoelastic effects.

## 2.1 Fundamental relations

We consider an isotropic material whose elastic stress tensor $\sigma$ can be expressed as a function of the Finger strain tensor **B**, defined as $\mathbf{F}\,\mathbf{F}^T$ where **F** is the displacement gradient tensor (Macosko 1994; Mal & Singh 1991). Since the principal axes of **B** are the only preferential directions of the material, the principal axes of $\sigma$ must coincide with those of **B**. For a simple shear strain $\gamma$ in the $x_1$ direction, the diagonal frame differs from the one shown in Fig. 1 by a rotation $\theta$ around the $x_3$ axis. Diagonalizing the tensors **B** and $\sigma$ yields respectively the results: $2\cot(2\theta) = \gamma$ and $2\cot(2\theta) = N_1/\sigma_{12}$ (Larson 1988). Comparing these two expressions leads to the Poynting relation Eq. (1).

To discuss the specific case of aqueous foams, a nonlinear constitutive rheological law is required. We consider the foam as incompressible and assume that its rheological behavior is isotropic. Moreover, we model the foam as a hyperelastic material (Mal & Singh 1991): All mechanisms of dissipation are ignored and we suppose that the mechanical response can be derived from a well-defined elastic strain energy density. Under these conditions, general theorems of nonlinear continuum mechanics show that the stress tensor can be expressed as follows (Macosko 1994):

$$\boldsymbol{\sigma} = -p\,\boldsymbol{I} + \beta_1\,\boldsymbol{B} + \beta_{-1}\,\boldsymbol{B}^{-1} \qquad (2)$$

Since we consider an incompressible material, the pressure p may be regarded as an arbitrary parameter. The material functions $\beta_1$ and $\beta_{-1}$ can in general depend on the invariants of **B**. Replacing them by their respective values in the limit of vanishing strain yields an equation of the Mooney-Rivlin form which describes approximately the nonlinear elastic response of many incompressible soft elastic materials (Macosko 1994). In a recent publication, we have presented theoretical arguments showing that if dissipation effects are neglected, such a law can indeed be expected to provide a good description of aqueous foams and concentrated emulsions, using $\beta_1 = G/7$ and $\beta_{-1} = -6G/7$ where G is the linear static shear modulus (Höhler et al 2004). This prediction agrees with recent numerical simulations and previous analytical work (Doi & Ohta 1991; Kraynik & Reinelt 2004; Kraynik et al 2000; Larson 1997). To conclude this paragraph, let us note that for a simple shear γ as illustrated in figure 1, the Mooney-Rivlin model for foams yields the following stress response:

$$\sigma_{12} = G\gamma \qquad N_1 = G\gamma^2 \qquad (3)$$

Moreover, the elastic strain energy density may be expressed as (Höhler et al 2004).

$$W = \frac{G}{14}((I_B - 3) + 6(II_B - 3)) \qquad (4)$$

where $I_B$ and $II_B$ are respectively the first and second invariants of the Finger strain tensor(Macosko 1994).

## 2.2 Quasi-static elastic response and the influence of trapped strains

As pointed out in the introduction, it is questionable to what extent relation (1) may apply to yield strain materials such as foams, where the flow history leads to trapped internal strains. Strains can be trapped if the boundary conditions connecting neighboring parts of the sample do not allow all regions of the material to be simultaneously in a stress free state. Such strains are common in solids or solid-like complex fluids where they arise in the neighborhood of regions where plastic flow occurs (Eshelby 1957; Picard et al 2004). In crystals, trapped strains exist in the neighborhood of dislocations (Kosevich et al 1986).

As a simple example illustrating how normal stresses can be modified if trapped strains are superposed to the shear strain applied by a rheometer, let us consider a static homogeneous trapped strain superposed to an applied static homogeneous strain, both chosen so small that no plastic flow occurs. Trapped and imposed strains are respectively described by displacement gradients denoted $\mathbf{F}_t$ and $\mathbf{F}_s$:

$$\mathbf{F}_s = \begin{pmatrix} 1 & \gamma & 0 \\ 0 & 1 & 0 \\ 0 & 0 & 1 \end{pmatrix} \quad \mathbf{F}_t = \mathbf{R}^T \begin{pmatrix} \Phi & 0 & 0 \\ 0 & \Psi & 0 \\ 0 & 0 & \Phi^{-1}\Psi^{-1} \end{pmatrix} \mathbf{R} \qquad (5)$$

**R** is a rotation matrix and $\Phi$ and $\Psi$ are constants describing the principal stretches in the diagonal frame of $\mathbf{F}_t$. It should be noted that $\mathbf{F}_t$ represents the most general isochoric homogeneous deformation. The combined effect of the imposed and trapped deformations is given by the displacement gradient tensor $\mathbf{F} = \mathbf{F}_s \mathbf{F}_t$ and the corresponding Finger strain tensor **B**:

$$\mathbf{B} = (\mathbf{F}_s \mathbf{F}_t)(\mathbf{F}_s \mathbf{F}_t)^T \qquad (6)$$

By inserting Eq.(6) into Eq.(2) with $\beta_1 = G/7$ and $\beta_{-1} = -6G/7$ we predict the dependence of the normal stress difference $N_1$ on the applied strain $\gamma$ and on the trapped strain, described by the stretch parameters $\Phi$ and $\Psi$ as well as the three Euler angles defining the rotation **R**. As a scalar measure of the amount of trapped strain, we consider the corresponding elastic energy density $W_t$ which can be calculated using Eq. (4). We expect the energy density at the yield shear strain, denoted $W_y$, to be an upper bound for the possible values of $W_t$. The experimental data presented in detail in section 5 show an onset of yielding for a shear strain of the order of 0.1, implying $W_y \cong 5 \cdot 10^{-3} G$. Fig. 2 gives some examples, showing that the nonlinear normal stress response to an applied shear can be strongly modified by trapped strains, even if their elastic energy density is much smaller than $W_y$. Trapped strains can add an offset to the first normal stress difference or to the applied strain. To a lesser extent, they can also change the second derivative of $N_1$ with respect to $\gamma$. Such phenomena have indeed been encountered in previous unsuccessful attempts to measure the normal stress response of foams reported in the literature (Khan et al 1988). In view of these findings, the question arises whether the influence of trapped strains can be separated experimentally from the intrinsic rheological behavior. We will show that the mentioned offsets can be discarded efficiently by studying the normal stress amplitude in response to an oscillating imposed shear strain and that such measurements give indeed access to features of the intrinsic rheological behavior, such as the Poynting relation.

In the context of nonlinear rheology an imposed oscillatory strain $\gamma(t) = \gamma_0 \mathrm{Re}[\exp(i\omega t)]$ induces a spectrum of stress oscillations at frequencies that are integer multiples of $\omega$. Often, the fundamental component of this spectrum is dominant, so that a generalized dynamic shear modulus taking into account only this part of the stress response is a useful measure of the viscoelastic response:

$$G^*(\omega, \gamma_o) = \frac{2}{T\gamma_o} \int_0^T \sigma(t)\, e^{-i\omega t}\, dt \qquad (7)$$

$T$ is the period of the oscillation. Eq. (7) leads in the linear regime to a complex shear modulus G* such that $\sigma(t) = \mathrm{Re}[G^* \gamma_0 \exp(i\omega t)]$. The first normal stress difference induced by an oscillating imposed shear strain also generally presents a spectrum of harmonics. For an isotropic material, the lowest frequency of this spectrum is $2\omega$, and we describe the corresponding component by the coefficient $N_1^*(2\omega, \gamma_0)$:

$$N_1^*(2\omega, \gamma_0) = \frac{2}{T} \int_0^T N_1(t)\, e^{-2i\omega t}\, dt \qquad (8)$$

If the Poynting relation holds, the following dimensionless ratio, denoted P, must be equal to 1:

$$P \equiv \left| \frac{2\, N_1^*(2\omega, \gamma_o)}{\gamma_o^2\, G^*(\omega, \gamma_o)} \right| = 1 \qquad (9)$$

Let us note that $N_1^*(2\omega, \gamma_0)$ varies quadratically with $\gamma_0$ according to the Mooney-Rivlin model for foams (see Eq. (3)), so that P is indeed predicted to be independent of the strain amplitude. This feature is conserved in the presence of trapped strains. To study the quantitative influence of arbitrary trapped strains on the validity of Eq. (9), rotation matrices

**R** were generated using random Euler angles and the stretch parameters $\Phi$ and $\Psi$ were randomly chosen in the range 0.85 - 1.15. The inset in Fig. 3 shows that this procedure allows to sample all of the parameter space where the elastic energy density satisfies the criterion $W_t < W_y$. In the principal graph of Fig. 3, P is plotted as a function of the trapped energy density, for each value of **R**, $\Phi$ and $\Psi$. The envelope of these points constitutes a bound on the possible influence of trapped strains on P. We conclude that within the limits of validity of the Mooney-Rivlin model for foams, the second harmonic normal stress response gives robust information about the constitutive law since the parameter P, allowing to assess the validity of the Poynting relation, is modified by no more than 20% in the presence of arbitrary trapped strains. Our analysis is meant to clarify only the principle of the oscillating strain measurements of normal stresses. In practice, trapped stresses may be of a well defined direction imposed by the flow history, or they may be partially relaxed due to creep flow, leading to bounds that can be more restrictive than those presented in figure 3. It would be interesting to extend our analysis to spatially heterogeneous trapped stresses, but such a calculation would require nonlinear elastic homogenization techniques which are beyond the scope of this paper.

## 2.3 A Nonlinear Viscoelastic Constitutive Model

In viscoelastic materials, the stress is not only a function of the instantaneous strain, but also of the preceding strain history, weighted by a memory function. As a consequence, the diagonal frames of **B** and **σ** will not coincide for oscillating shear strains and an exact analogue of the Poynting effect cannot be expected in general. To discuss the specific case of foams, we construct a minimal nonlinear viscoelastic model: We start from a linear viscoelastic Maxwell model, whose characteristic time scale $\tau$ is governed by the rate of coarsening induced bubble rearrangements. This approach is motivated by previous experiments and simulations described in the introduction. We consider this model in its integral form (Larson 1988) and replace the term describing the linear elastic stress by the expression predicted using the Mooney-Rivlin model (Höhler et al 2004):

$$\boldsymbol{\sigma}(t) = -p\mathbf{I} + \int_{-\infty}^{t} \frac{e^{-\frac{t-t'}{\tau}}}{\tau} \left( \beta_{-1} \mathbf{B}^{-1}(t,t') + \beta_{1} \mathbf{B}(t,t') \right) dt' \qquad (10)$$

$\mathbf{B}(t, t')$ and $\mathbf{B}^{-1}(t, t')$ describe the strain evolution over the time interval [t, t']. It is interesting to note that the dominant term in Eq. (10), which contains the coefficient $\beta_{1}$, corresponds to the lower convected Maxwell model (Larson 1988). From a formal point of view, we note that Eq. (10) is a member of time-strain factorized Kaye-BKZ class of constitutive equations well known in polymer rheology (Larson 1988). The hypothesis that it is possible to factorize the integrand into two terms respectively depending on time and strain only is non-trivial. In our case, it amounts to neglecting strain induced rearrangements as a source of viscoelastic relaxation. As the yield strain is approached, this assumption breaks down as will be discussed in the following section. Eq. (10) predicts the following stress response to an oscillating shear strain:

$$G^{*}(\omega, \gamma_0) = \frac{G \, i\omega\tau}{1 + i\omega\tau} \qquad N_1^{*}(2\omega, \gamma_0) = \frac{G \, \gamma_0^{2} \omega^{2} \tau^{2}}{-1 - 3i\omega\tau + 2\omega^{2}\tau^{2}} \qquad (11)$$

Let us compare these results to the Poynting relation in the form of Eq. (9), expressed in terms of the parameter P. For the viscoelastic rheological behavior described by Eq. (11), P is indeed a robust quantity since it is independent of the elastic modulus G and the strain amplitude $\gamma_0$. P depends only on the Deborah number De = $\omega\tau$, as illustrated in Fig. 4. As expected, P tends towards 1 in the limit of large De, corresponding to elastic behavior. Remarkably, the Poynting relation in the form of Eq. (9) continues to be a good approximation in the presence of strong dissipation: We obtain $0.88 \leq P \leq 1$ down to a Deborah number of 1, where viscous and elastic stresses are of equal amplitude. P goes to zero in the limit of small Deborah numbers where viscous stresses become dominant. The constitutive equation (10) can be refined and applied in an extended range of frequencies if additional relaxation processes are included in the memory function. Indeed, such processes are known to exist in foams, and they have been evidenced by the creep experiments mentioned in the introduction. We have checked that for a memory function constructed from such data, the relation P $\cong$ 1 still applies for Deborah numbers much larger than one.

## 3. Materials

The foam samples were generated by injecting a gas and a polymer-surfactant-based aqueous solution into a column filled with glass beads, as described elsewhere[21]. The foaming formulation contained sodium α-olefin sulfonate (AOK, Witco Chemicals), Polyethylene-oxide ($M_w = 3 \times 10^5$ g mol$^{-1}$, Aldrich) and dodecanol (Aldrich), (concentrations: 1.5% g/g, 0.2% g/g and 0.4% g/g respectively). The injected gas was pure nitrogen. To characterize the AOK solution, its surface tension was measured by the Wilhelmy plate method, and found to be equal to 22.6 mN/m. Its viscosity was determined using an Ostwald capillary viscometer, and found to be equal to 1.9 cP. The measured gas volume fraction of all the foam samples is (97.0 ± 0.3) %. Immediately after its production, the foam was injected into the measuring cell of the rheometer. All rheological measurements were performed 20 min after the injection. At this time, the average bubble diameter was 215 μm.

Gel samples were used to check the accuracy of our special purpose rheometer described in the following section. They consisted of a poly(vinyl) alcohol (PVA), cross-linked by sodium tetraborate decahydrate (Borax). We used a PVA of average molecular weight $M_w$ = 85000 -146000 g mol$^{-1}$, and of degree of saponification equal to 96%. Chemicals were bought from Aldrich and used as received. The gel was prepared by adding some 4.0%g/g Borax aqueous solution to a 4.0%g/g PVA aqueous solution with rapid and vigorous stirring until the solution had gellified. The proportions of both solutions were adjusted so that the final mass ratio of PVA to Borax in the gel was equal to 24.5. All material characterizations were carried out at a temperature of (21 ± 1) °C.

## 4. Experimental Methods

To study the first normal stress response to an applied oscillating shear strain, we use a rheometer equipped with a cone and plate cell: The sample is confined between a horizontal plate and a cone whose axis is perpendicular to the plate. The first normal stress difference can be deduced from the total vertical thrust F on the plate using the relation $N_1 = 2F/(\pi r^2)$ where r is the cone radius (Macosko 1994). To study the bulk properties of foams, it is necessary that the gap between the plate and the cone be much larger than the bubble size in as large a fraction of the sample as possible. We have therefore used cone angles β of either 10° or 15° and a cone diameter of 120 mm. For these two values of β, the gap width is larger than 10 bubble diameters in respectively 96% and 98% of the plate area covered by the foam. We recall that in the cone plate geometry, the ratio of the minimum value of the shear stress (on the plate surface) to its maximum value (on the cone surface) varies as $\sin^2(90 - \beta)$ [20]. For β = 10° and β = 15°, the heterogeneity of the applied stress in the sample remains below 3% and 7% respectively. Another potential source of stress heterogeneity are deviations of the shape of the free boundary of the foam samples which is in our experiments cylindrical rather than spherical, as supposed in the standard analysis of the cone plate geometry (Macosko 1994). This effect is expected to be a function of cone angle. As discussed in the following section, we have compared data obtained at both cone angles to check such artifacts. A video camera was used to monitor the bubble motion on the free surfaces of the sample to check the absence of shear banding and wall slip. To avoid this latter effect, all cone and plate surfaces in contact with the samples were grooved radially.

For soft materials such as aqueous foams and gels, normal stress measurements over an extended range of strain amplitudes below the yield strain require a rheometer whose sensitivity is well beyond that of existing commercial devices. We have therefore constructed a special purpose rheometer adapted for cone plate geometries of very large diameter. This device uses a lock-in amplifier to detect the very small second harmonic normal forces induced by an oscillating shear strain accurately. To test our special purpose rheometer, we have compared data measured for a PVA/Borax gel, using either this device or a Bohlin CVOR150 rheometer with a standard cone-plate geometry (cone diameter 40 mm, $\beta = 4°$). The shear modulus of the gel was similar to that of the foam, but since this gel fractures only at strain amplitudes much larger than the yield strain of foams, normal stresses much stronger than in foams could be produced, sufficient to be detected in the commercial instrument. Agreement between the data obtained in these tests with both devices is excellent. Moreover, the special purpose rheometer allows normal forces to be measured reliably with a sensitivity that is 100 times better than the one of research grade commercial instruments. To determine the complex shear modulus, we used the Bohlin rheometer equipped with a cylindrical Couette geometry (inner radius 21 mm, gap width 4 mm, cylinder height 48 mm). To avoid wall slip, the surfaces of the cylinders were grooved.

Since foams are yield stress fluids and conserve a memory of their strain history, care was taken to control as much as possible the flow of the sample upon its installation in the rheometer. In the cone plate geometry, the foam was first deposited on the plate, and then the cone was lowered at a speed of 0.1 mm/s, inducing a squeeze flow. We then applied an oscillating strain of 1 Hz with an initial amplitude of 0.45. This amplitude was reduced to zero in five steps, lasting 1 minute each. Such a symmetrical preshearing was used, since the normal stress response critically depends on the isotropy of the sample structure (cf. section 2.1). As described in the following, the preshearing has indeed an influence on the observed normal stress response. In the shear modulus measurements carried out using the cylindrical Couette geometry, the effect of such preshearing was found to be negligible. All rheological measurements were performed at a frequency of 1 Hz and at a temperature of (21 ± 1) °C. The air in contact with the sample was saturated with humidity to avoid evaporation.

## 5. Results and Discussion

Figure 5 gives an overview of the foam normal stress response as a function of strain amplitude $\gamma_0$. Before the normal stress measurement, each sample was subjected to the preshearing procedure described in the experimental section. The data are shown only up to a strain amplitude of 0.5, since beyond this value, observation of the free lateral surface of the sample revealed the onset of strain localization. A quadratic scaling of the normal stress amplitude with $\gamma_0$ appears over a large range of strain amplitudes. Figure 6 shows the real and imaginary parts of the complex shear modulus, G' and G", as a function of strain amplitude. The yield strain appears as a drop of G' and a maximum of G", as in previous results obtained at low frequencies for aqueous foams of similar gas volume fraction (Saint-Jalmes & Durian 1999). Up to a strain amplitude of 0.1, G' and G" do not present any strong evolution with $\gamma_0$ in agreement with our nonlinear viscoelastic model.

Figure 7 shows the ratio P defined by Eq.(9) as a function of strain amplitude $\gamma_0$. For a cone angle of 15° and 0.04 < $\gamma_0$ < 0.1 where our viscoelastic model is expected to apply, P remains in the range 0.9 ± 0.1. This result agrees with the predicted value 1 for a Deborah number De >> 1, estimated as G'/G" using the data shown in figure 6. At strain amplitudes lower than 0.04, the scatter of the data for β = 15° increases, but the average ratio does not evolve significantly. For β = 10° and $\gamma_0$ < 0.1, the scatter of the data is very strong and the average value of P is significantly larger than 1. It is interesting to note that a less pronounced but similar effect is obtained in experiments using the 15° cone if no preshearing is applied to the foam before the normal stress measurement. These data suggest that the deviations from the Poynting law are linked to the plastic squeeze flow upon sample installation since it is strongest for the smallest cone angles, and since preshearing is expected to erase flow history at least partially. Oscillatory preshearing in a single direction is manifestly not always sufficient to erase the rheological memory of flow history. The violation of the Poynting relation at low strains cannot be explained in terms of fixed trapped strains in an intrinsically isotropic hyperelastic material, since our model presented in section 2.2 predicts P to be independent of strain amplitude. Simulations using the Surface evolver software could help to identify the underlying mechanism on the bubble scale.

At strain amplitudes $0.1 \leq \gamma_0 \leq 0.5$, the results obtained for cone angles of 10° and 15° coincide, excluding the cone angle dependent artifacts mentioned in the experimental section. In this regime, the Poynting relation expressed as P = 1 applies to a good approximation to the studied aqueous foams, but it is only for $\gamma_0 \leq 0.1$ that this feature is predicted and explained by our viscoelastic model. Indeed, for $\gamma_0 \gg 0.1$, strain induced irreversible bubble rearrangements are expected to be the origin of the observed drop of G' and maximum of G'' with strain amplitude and our analysis in the framework of hyperelastic or nonlinear viscoelastic models does not apply to such plastic flow. Most existing models of yielding are not able to predict normal stresses since they describe stress and strain schematically as scalar quantities (Derec et al 2001; Hébraud & Lequeux 1998; Sollich et al 1997). Only recently, a tensorial model has been published, which could in principle be compared to our results (Cates & Sollich 2004). However, predicting the rheological response to large amplitude oscillating shear in this framework requires non-trivial numerical calculations that are beyond the scope of the present paper.

## 6. Conclusion

In this paper, we report the first experimental study of shear induced normal stress differences in solid-like aqueous foams. It is shown experimentally and explained theoretically that measuring the first normal stress response to oscillating shear gives more robust results than static normal stress measurements. Furthermore, we present a physically motivated nonlinear viscoelastic model for foam, predicting a relation between normal stress and shear stress oscillation amplitudes. This analog of the Poynting relation is in good agreement with our data for dry foams if the experimental geometry is chosen such that squeeze flow upon sample installation in the rheometer is weak and if trapped strains are minimized using a specific preshearing procedure. Even at strain amplitudes close to the yield strain where plastic flow sets in, the Poynting relation still holds to a good approximation. This unexpected feature is not explained by existing models of foam rheology. We expect that these findings will help to construct and to test future constitutive models explaining the nonlinear rheological behavior of aqueous foams and concentrated emulsions, and possibly a much wider class of complex yield strain fluids.

## Acknowledgements

We would like to acknowledge stimulating discussions with Michael Cates, Peter Sollich, Denis Weaire, Andrew Kraynik, Qi-Chang He and Guy Bonnet, and we thank David Hautemayou for his efficient technical help. This work was supported by the MENRT through the EA 2179 and by the CNRS via the UMR 8108.

**Figure Captions**

Fig 1: Illustration of the notation used to describe the first and second normal stress differences, induced by a simple shear strain in the $x_1$ direction.

Fig 2: The first normal stress difference, normalized by the linear static shear modulus G, is plotted as a function of the applied shear strain $\gamma$, as predicted by the Mooney-Rivlin law for foams presented in section 2.1. The thick full line is predicted in the absence trapped strain. The other two curves are examples illustrating how this result can be modified by trapped strains, chosen well below the yield strain: Their elastic energy density $W_t$ is 0.2 $W_y$, with $W_y = 0.005$ G. The dashed line corresponds to the stretch parameters $\Phi = 1.046$, $\psi = 1$, and a rotation of $\pi/4$ around the $x_3$ axis. The thin full line is obtained for $\Phi = 1$, $\psi = 0.955$ and a rotation of -0.75 rad around the $x_3$ axis.

Fig 3: The ratio P, allowing to assess the validity of the Poynting relation and defined in Eq. (9), is plotted versus the trapped elastic energy density $W_t$, normalized by the elastic energy density at the yield strain $W_y$. Each point corresponds to a strain randomly oriented with respect to the applied shear, with stretch parameters randomly chosen in the intervals $0.85 < \Phi < 1.15$ and $0.85 < \psi < 1.15$. The insert shows the locus of stretch parameters where yielding sets in and where $W_t = W_y$. The domain enclosed by this locus corresponds to all possible configurations where the mechanical response is elastic and $W_t < W_y$. This graph indicates that the investigated range of $\Phi$ and $\psi$ covers indeed all possible trapped strains.

**Fig 4:** The ratio P, defined in Eq. (9), is plotted versus the Deborah number De. This result is predicted by the viscoelastic model leading to Eq. (11).

**Fig 5:** Measured second harmonic normal stress oscillation amplitude as defined in Eq. (8) versus applied shear strain amplitude. Open and filled symbols correspond to measurements using a cone angle of 10° and 15° respectively. In each case, the results of three independent experiments are shown. The straight line corresponds to a quadratic law.

**Fig 6:** The measured real and imaginary parts of the complex shear modulus, as defined in Eq. (7), are plotted as a function of strain amplitude. The lines are guides to the eye.

**Fig 7:** The measured ratio P, defined in Eq. (9), is shown as a function of strain amplitude. Open and filled symbols correspond to measurements using a cone angle of 10° and 15° respectively. In each case, the results of three independent experiments are shown. The horizontal line indicates the value of P predicted by the nonlinear viscoelastic model presented in section 2.3 for a Deborah number De >> 1.

**Figure 1**

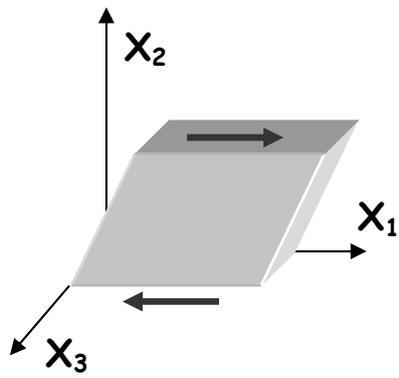

$N_1 = \sigma_{11} - \sigma_{22}$
$N_2 = \sigma_{22} - \sigma_{33}$

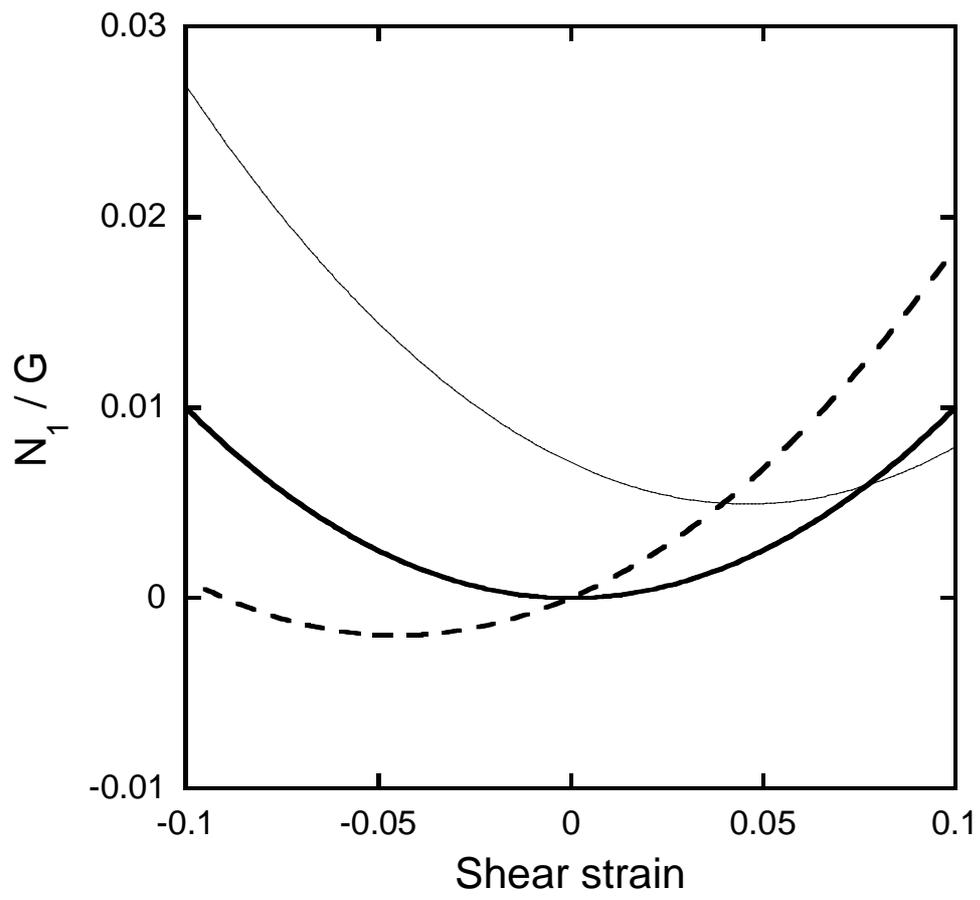

**Figure 2**

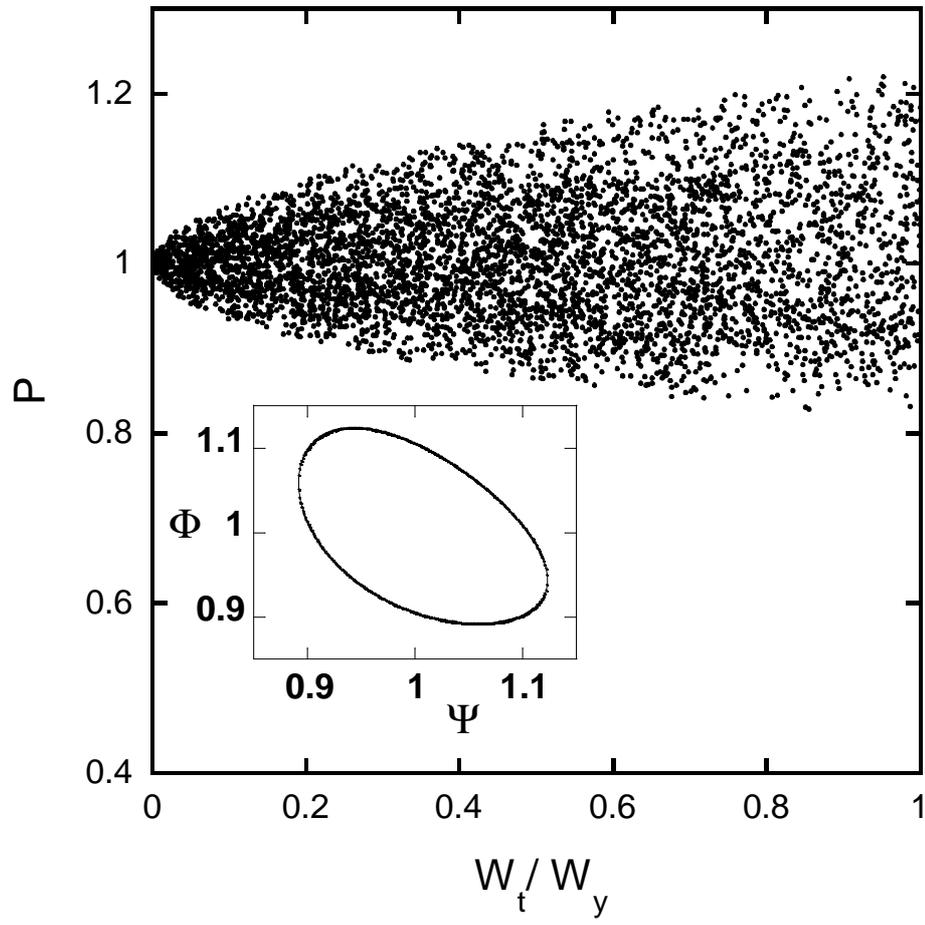

**Figure 3**

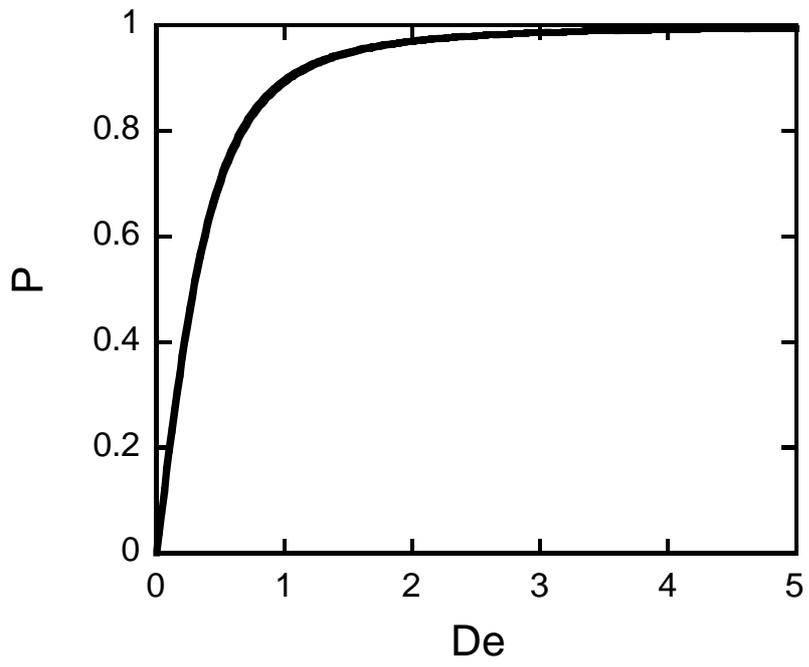

**Figure 4**

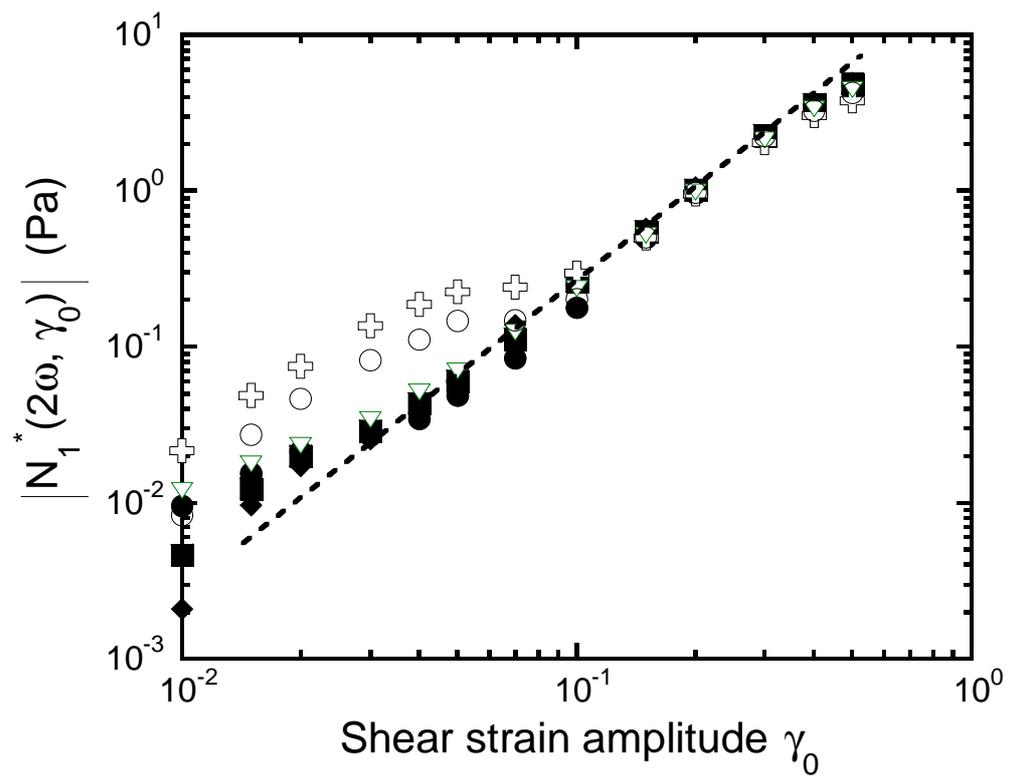

**Figure 5**

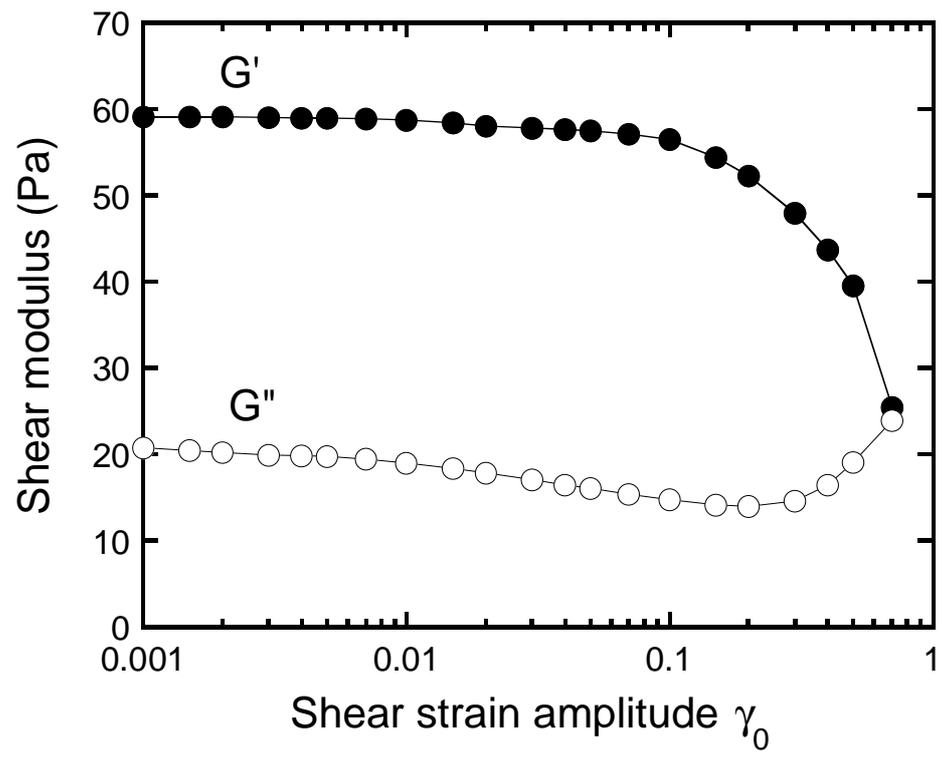

**Figure 6**

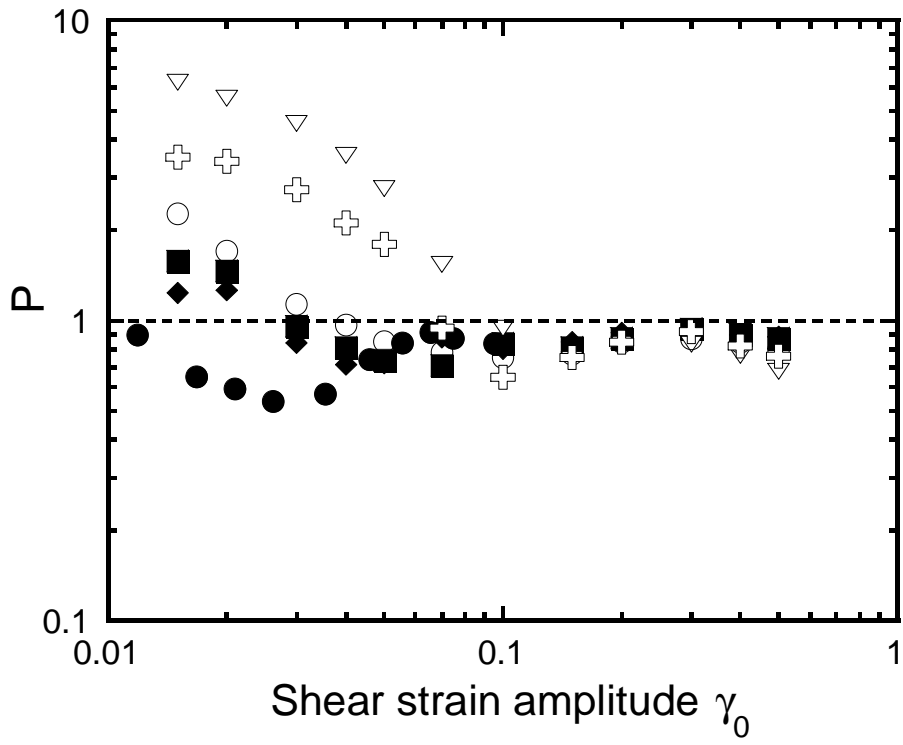

**Figure 7**